\documentclass[preprint,showpacs,preprintnumbers,amsmath,amssymb,a4paper,showkeys]{revtex4}

\usepackage{graphicx}	
\usepackage{dcolumn}	
\usepackage{bm}		
\usepackage{epsfig}
\usepackage{fancyhdr}

\begin{document}

\newcommand{\ud}{\mathrm{d}}		
\newcommand{\EPS}{\varepsilon}          
\newcommand{\kommentti}[1]{}		
\newcommand{\V}[1]{\mathbf{#1}}		
\newcommand{\iim}{\mathrm{i}}	
\newcommand{\Op}[1]{\mathcal{#1}} 		
\newcommand{\ket}[1]{| #1 \rangle}		
\newcommand{\bra}[1]{\langle #1 |}		
\newcommand{\me}[3]{\bra{#1} #2 \ket{#3}}	
\newcommand{\cre}[2]{\hat{#1}_{#2}^{\dagger}}	
\newcommand{\ann}[2]{\hat{#1}_{#2}}		

\title{Cavity photon counting: ab-initio derivation of the quantum jump superoperators and comparison of the existing models}

\author{Teppo~H\"ayrynen}	\email{tihayryn@lce.hut.fi}
\author{Jani~Oksanen}	        
\author{Jukka~Tulkki}		
 \affiliation{
 Department of Biomedical Engineering and Computational Science, 
 Helsinki University~of~Technology,P.O.~Box~9203, FIN-02015~HUT, Finland
 }

\date{\today}

\begin{abstract}

Time development of electromagnetic fields in closed cavities under continuous detection
of photons continues to be a subject of confusing controversy.
Recently Dodonov \textit{et al.} [Phys. Rev. A, \textbf{75}, 013806, 2007]
argued that their model of quantum superoperators (E model)
invalidates some of the predictions of the previously introduced photon counting
model of Srinivas and Davies [J. Mod. Optic. \textbf{28}, 981, 1981] (SD model).
Both the SD and the E models are based on two postulated quantum jump superoperators:
(1) the one-count operator corresponding to the absorption of a single photon
and (2) the no-count operator.
In this work we develop a stochastic difference equation that describes the dissipative coupling of the cavity field and the detector.
The  difference equation is based on non-perturbative treatment of the cavity-detector coupling. 
In spite of being non-integrable due to the coupling of the detector with an external reservoir it
can be used to derive the exact forms of the quantum jump superoperators.
When applied to a particular photon counting measurement our theory 
gives predictions identical with those of the SD model
which should be considered a \textit{non-perturbative} and \textit{ab-initio} result.
It is pointed out that available   experimental results 
coincide with the results given by the ab-initio SD model.
We summarize some of the key characteristics of cavity fields
and photon counting processes to demonstrate that the results 
given by the SD model are consistent with the principles of
quantum mechanics while those given by the E model are not.

\end{abstract}

%
%
\pacs{42.50.Ar, 42.50.Lc, 03.65.Ta}

\keywords{quantum optics, optical cavity,  dissipative system, 
continuous photon detection, photon counting, SD model, E model}

\maketitle


\section{Introduction}

The theory of  cavity photon counting  and in particular 
the expected evolution of the cavity field after detection of a 
sequence of photons continues to be a subject of confusing controversy 
as shown in a recent report by Dodonov \textit{et al.} \cite{DMD07}. 
So far no experimental data on the evolution of the
cavity fields are available to resolve the controversy .
The experiments themselves should be
feasible at present though. 
This has prompted us to study the relation between the photon counting models.
In this work we consider a  system formed by the cavity
field, the detector, and a (infinite) reservoir that is coupled
to the detector.

The first quantum mechanical models of photon counting
were formulated for an experiment where a light
beam entered the detector and the unabsorbed photons were able to escape \cite{KK, M81}.
Those models were based on the quantum version of the classical Mandel's formula \cite{SD}.
In these models there was thus no need to consider the measurement back action.
However, in the case of a cavity field the time integrated effect of the measurement 
on the electromagnetic field will be large 
and therefore the measurement back action cannot be ignored.
We will show that the analysis of photon detection measurements
can be carried out starting from the interaction Hamiltonian of the 
field and the detector.
It is pointed out that for the photon counting analysis there is no need of introducing new postulates 
beyond the well established theoretical foundations of cavity quantum electrodynamics.
It turns out that our non-perturbative approach gives results
which agree with the SD model \cite{SD} 
while the E model \cite{DMD07, OMD, DMD05, DMD06}
is shown to be inconsistent. 
Furthermore, we show that the measured data of the second order 
coherence degrees of thermal and coherent fields \cite{LLZ}
agree with the SD model.

\section{Photon counting based on open system master equation}
\label{QED_mod}

The SD and E photon counting models were originally introduced using phenomenological 
arguments or postulates. The first rigorous approach to cavity photon counting was
formulated by Imato \textit{et al.} \cite{IUO}. They used a homogeneous atomic
beam as a tool for including the dissipative terms in the Jaynes-Cummings Hamiltonian.
The dissipation of the field was modeled using a sequence of infinitesimal
perturbations caused by the atoms.

In this work we show that the atomic beam model is not needed
for derivation of the theory of cavity photon counting.
Instead we assume that the whole system consists of three
parts: the cavity, the detector and a (infinite) reservoir.
The details of the coupling between detector and reservoir
are unknown but it is assumed that this coupling
is strong enough so that the detector is most of the time in its
ground state $\ket{g}$. The time the detector will spend in its
lowest excited state $\ket{e}$ is so short that the probability
of a cavity photon to interact with the detector in the excited
state is vanishingly small. 
Accordingly in any short time interval  only absorption of one photon
or no photons are possible i.e. the  one-count or no-count events, respectively. 
The photoabsorption processes are Markovian processes i.e. the
system is memoryless and the absorption probability depends only
on the current state of the system, not on the past states. 
Furthermore, we assume that the detector doesn't return to
the ground state by emitting the photon back to the cavity.

The optical cavity is assumed ideal and thus the only dissipative mechanism included
is the absorption of photons by the detector. Furthermore, because the
detector is assumed to return back to the ground state immediately after the 
absorption of a photon we do not consider the detector saturation effects. 
If we assume that the photons escape from the cavity 
to a detector which is, for example, a photomultiplier tube, 
the photons would dissipate from the cavity to the photomultiplier tube even if
the tube was be momentarily switched off. Thus we can neglect the dead time of the detector.

We assume that the interaction of the field and the detector is described
by the Jaynes-Cummings Hamiltonian in the rotating wave approximation
\begin{equation}
\hat{H}_I = \hbar\Omega \Big( \cre{a}{}\ket{g}\bra{e} + \ann{a}{}\ket{e}\bra{g} \Big). \label{H_int} 
\end{equation}
Furthermore, the field Hamiltonian is $\hbar\omega\cre{a}{}\ann{a}{}$, the 
detector Hamiltonian is $\hbar\omega\ket{e}\bra{e}$, and the initial detector-field
density operator is
\begin{equation}
\hat{\rho}_I(0) = \sum_{n,n'=0}^{\infty}p_{n,n'} \ket{n,g}\bra{n',g}. \label{rho_0}
\end{equation}

We next find an expression for the differential change of the 
detector-field density operator 
$\ud \hat{\rho}_I$ = $\hat{\rho}_I(\ud t) - \hat{\rho}_I(0)$.
At the starting point of the interval $[0, \ud t]$ the detector
is in the ground state and the probability of the
detector being excited to $\ket{e}$ is small. Therefore during
$[0, \ud t]$ the coupling of the detector with the reservoir
can be neglected since the reservoir only causes relaxation 
of the detector from the excited state to the ground state.
Thus the field detector subsystem can be considered closed
within $[0, \ud t]$. The differential change of the density operator
can then be obtained \textit{non-perturbatively} using the Taylor series
\begin{eqnarray}
\hat{\rho}_I(dt)-\hat{\rho}_I(0) &=&  \left. \frac{\ud \hat{\rho}_I(t)}{\ud t}\right|_{t=0} \ud t 
  + \frac{1}{2}\left. \frac{\ud^2 \hat{\rho}_I(t)}{\ud t^2}\right|_{t=0} (\ud t)^2 + \dots.
\end{eqnarray}
As shown in detail in the Appendix \ref{sec:qed_der},
the Taylor expansion gives (see Appendix \ref{sec:qed_der}.)
\begin{equation}
\frac{\hat{\rho}_I(\ud t)- \hat{\rho}_I(0)}{\ud t} = -\frac{\iim}{\hbar}[\hat{H}_I, \hat{\rho}_I(0)] + 
  \frac{\ud t}{2\hbar^2}\left( 2\hat{H}_I\hat{\rho}_I(0)\hat{H}_I - \{ \hat{H}_I\hat{H}_I,\hat{\rho}_I(0) \} \right). \label{lf_ime}
\end{equation}
Note that this result is exact in the limit of infinitesimal $\ud t$.  
From equations (\ref{H_int}) and (\ref{rho_0}) we see that
$\hat{H}_I\hat{\rho}_I(0)$ = $\hbar\Omega\sum_{n,n'} p_{n,n'}\sqrt{n}\ket{n-1,e}\bra{n',g}$ = $\hbar\Omega\ann{a}{}\ket{e}\bra{g}\hat{\rho}_I(0)$, that
$\hat{H}_I\hat{H}_I\hat{\rho}_I(0)$ = $(\hbar\Omega)^2\sum_{n,n'} p_{n,n'} n \ket{n,g}\bra{n',g}$ = $(\hbar\Omega)^2\cre{a}{}\ann{a}{}\hat{\rho}_I(0)$, and that
$\hat{H}_I\hat{\rho}_I(0)\hat{H}_I$ = $(\hbar\Omega)^2\sum_{n,n'} p_{n,n'} \sqrt{nn'} \ket{n-1,e}\bra{n'-1,e}$ 
= $(\hbar\Omega)^2\ann{a}{}\ket{e}\bra{g}\hat{\rho}_I(0)\ket{g}\bra{e}\cre{a}{}$.
Using these  relations and replacing the initial time $t=0$ with $t$, equation (\ref{lf_ime}) can be written as
\begin{eqnarray}
	\nonumber
\frac{\ud \hat{\rho}_I(t)}{\ud t} &=& \frac{\hat{\rho}_I(t+\ud t)- \hat{\rho}_I(t)}{\ud t}  \\
	\nonumber
&=&  -\iim\Omega\sum_{n,n'} p_{n,n'}(t)\bigg(\sqrt{n}\ket{n-1,e}\bra{n',g}-\sqrt{n'}\ket{n,g}\bra{n'-1,e} \bigg) \\
&& +  \frac{\Omega^2 \ud t}{2}\bigg( 2\ann{a}{}\ket{e}\bra{g}\hat{\rho}_I(t)\ket{g}\bra{e}\cre{a}{} - \{ \cre{a}{}\ann{a}{},\hat{\rho}_I(t) \} \bigg). \label{lf_ime2}
\end{eqnarray}
Equation (\ref{lf_ime2}) can not be solved by simple integration since it has been assumed 
that in the beginning of each time interval $[t, t+\ud t]$ the detector is at ground state
and therefore the coupling between the the detector and the reservoir has been neglected.
With these assumptions equation (\ref{lf_ime2})
describes the field-detector system as an open quantum system dissipating its energy to
an infinite reservoir represented by the detector returning to the ground state 
infinitely fast. 
In the following we are interested in the evolution of the field density operator. 
This is obtained from equation (\ref{lf_ime2}) by calculating the trace over the detector states
$\hat{\rho}_f$ = $\me{g}{\hat{\rho}_I}{g}$ + $\me{e}{\hat{\rho}_I}{e}$. 
Setting $\ud \hat{\rho}_f = \hat{\rho}_f(t+\ud t) -\hat{\rho}_f(t) $
and assuming that the detector is at ground state at $t$,
the reduced density operator i.e. the density operator of the field can be written as
\begin{equation}
\frac{ \ud \hat{\rho}_f(t)}{\ud t}  =  \frac{\Omega^2 \ud t}{2}\bigg( 2\ann{a}{}\hat{\rho}_f(t)\cre{a}{} - \{ \cre{a}{}\ann{a}{},\hat{\rho}_f(t) \} \bigg), \label{red_lf_ime2}
\end{equation}
Note that in contrast to equation (\ref{lf_ime2}), equation (\ref{red_lf_ime2}) may be solved
by integration for our detector model, since it no longer includes the detector states.
Substituting $\hat{\rho}_f(t)$ = $\sum_{n,n'=0}^{\infty}p_{n,n'}(t)\ket{n}\bra{n'}$
into  equation (\ref{red_lf_ime2}) gives
\begin{eqnarray}
	\nonumber
\frac{ \ud \hat{\rho}_f(t)}{\ud t}  &=& 
\sum_{n,n'=0}^{\infty}\frac{\ud p_{n,n'}(t)}{\ud t}\ket{n}\bra{n'}  \\
&=&  \Omega^2 \ud t \bigg( \sum_{n,n'=0}^{\infty}\sqrt{nn'}p_{n,n'}(t)\ket{n-1}\bra{n'-1}-\sum_{n,n'=0}^{\infty}\frac{n+n'}{2}p_{n,n'}(t)\ket{n}\bra{n'}  \bigg). \label{lf_ime3}
\end{eqnarray}
In order to relate the constant $\Omega^2\ud t$ to the coupling strength of the 
field-detector we solve the differential equation
for the probabilities $p_{n,n}$ i.e. the master equation of the photon numbers.
Taking the diagonal matrix elements $\me{n}{\cdot}{n}$ of equation (\ref{lf_ime3}) and 
denoting $p_n = p_{n,n}$ gives
\begin{eqnarray}
	\nonumber
\frac{\ud p_n(t)}{\ud t} &=& \frac{\ud \me{n}{\hat{\rho}_f(t)}{n} }{\ud t} \\
&=& \Omega^2\ud t \big(  (n+1)p_{n+1} - n p_n  \big). \label{lf_ime4}
\end{eqnarray}
Thus, the detector only absorbs photons and do not emit them to the cavity. 
The constant $\lambda \equiv \Omega^2\ud t$ is the probability per unit time that a photon 
will be absorbed by the detector i.e. the rate of absorption.

The physical interpretation of $\Omega^2\ud t$ becomes particularly
transparent if one considers the absorption of a single photon by
the detector. Let the amplitude of the detector ground state 
+ one photon field state ($\ket{g,1}$) be given by $C_g(t)$ 
and the amplitude of the excited state ($\ket{e,0}$) by $C_e(t)$
with initial conditions $C_g(0)=1$ and $C_e(0)=0$, respectively.
The differential change of $C_e(t)$ can be then obtained from the
time dependent Schr\"odinger equation. It is given by
$C_e(\ud t)$ = $-\iim V_{e,g}  \ud t/\hbar$, 
where 
$V_{e,g}$ = $\me{e,0}{ \hat{H}_I }{g,1}$ = $\hbar\Omega$
is the dipole amplitude.
The probability that the photon has been absorbed is 
$|C_e(\ud t)|^2$ = $|V_{e,g}|^2 \ud t^2/\hbar^2 $. 
Thus the absorption probability per unit time is given by
$|C_e(\ud t)|^2/\ud t$ = $|V_{e,g}|^2 \ud t /\hbar^2 $ = $\Omega^2 \ud t$. 
This is in accordance with equation (\ref{lf_ime4}).

\section{One-count and no-count superoperators derived from master equation }

The density operator of the system after infinitesimal time $\ud t$ is obtained using equation (\ref{lf_ime2}) 
and relation $\hat{\rho}_I(t+\ud t)$ = $\hat{\rho}_I(t) + \frac{\ud \hat{\rho}_I(t)}{\ud t} \ud t$
which gives
\begin{eqnarray}
	\nonumber
\hat{\rho}_I(t+\ud t) &=& \hat{\rho}_I(t) + \bigg[-\iim\Omega\sum_{n,n'} p_{n,n'}(t)\bigg(\sqrt{n}\ket{n-1,e}\bra{n',g}-\sqrt{n'}\ket{n,g}\bra{n'-1,e} \bigg) \\
&& +  \frac{\Omega^2 \ud t}{2}\bigg( 2\ann{a}{}\ket{e}\bra{g}\hat{\rho}_I(t)\ket{g}\bra{e}\cre{a}{} - \{ \cre{a}{}\ann{a}{},\hat{\rho}_I(t) \} \bigg)\bigg] \ud t.
\end{eqnarray}
After measuring a photoabsorption process the detector is in the excited state 
$\ket{e}$ so the part of the field density operator corresponding to the one-count event is given by $\me{e}{\hat{\rho}_I(t+\ud t)}{e}$.  
If a photon is not absorbed the detector stays in the ground state
and the part of the field density operator corresponding to the no-count-event is given by
$\me{g}{\hat{\rho}_I(t+\ud t)}{g}$.
We obtain
\begin{eqnarray}
\me{e}{\hat{\rho}_I(t+\ud t)}{e} &=& \lambda  \ann{a}{}\hat{\rho}_f(t)\cre{a}{} \ud t \label{qed_1ce} \\
\me{g}{\hat{\rho}_I(t+\ud t)}{g} &=& \hat{\rho}_f(t) - \frac{\lambda }{2}\left( \cre{a}{}\ann{a}{}\hat{\rho}_f(t) + \hat{\rho}_f(t) \cre{a}{}\ann{a}{} \right) \ud t. \label{qed_nce}
\end{eqnarray}
Therefore, after the sudden absorption of a single photon
the one-count operator is simply given by equation (\ref{qed_1ce})
\begin{equation}
\hat{J}\hat{\rho}_f(t) = \lambda  \ann{a}{}\hat{\rho}_f(t)\cre{a}{}. \label{qed_1co}
\end{equation}

In contrast to the abrupt one-count process the no-count process
changes the field density operator in a continuous manner.
Therefore it is necessary to calculate the no-count operator in a finite time interval $[t,t+\tau]$
between two absorption events.
The density operator is changed by the no-count event 
by the amount  $\ud\hat{\rho}_f(t)$ = $-\frac{\lambda }{2} \{ \cre{a}{}\ann{a}{},\hat{\rho}_f(t) \} \ud t$,
obtained  from equation (\ref{qed_nce}),
whose solution in time interval $[t,t+\tau]$ (here $\tau$ is not necessarily infinitesimal) gives the no-count operator
\begin{equation}
\hat{S}_{\tau}\hat{\rho}_f(t) = 
    e^{-\frac{\lambda}{2} \cre{a}{}\ann{a}{} \tau} \hat{\rho}_f(t) e^{-\frac{\lambda}{2} \cre{a}{}\ann{a}{} \tau}.
	\label{qed_nco}
\end{equation}

The probabilities of absorbing a photon and not absorbing a photon
are obtained from equations (\ref{qed_1co}) and (\ref{qed_nco}), respectively,
by calculating Trace$\{\hat{J}\hat{\rho}_I\}$/Trace$\{\hat{\rho}_I\}$
and Trace$\{\hat{S}_{\tau}\hat{\rho}_I\}$/Trace$\{\hat{\rho}_I\}$.
Equation (\ref{qed_1co})
gives the probability of the abrupt absorption process at $[t, t+\ud t]$  
\begin{equation}
P_{one-count}(t,t+\ud t) = \lambda \bar{n}(t) \ud t. \label{qed_1cp}
\end{equation}
The no-count probability at time interval $[t,t+\tau]$ is obtained calculating
the trace of (\ref{qed_nco}). Using the series expansion of the exponential terms
gives (see Appendix \ref{sec:ncp})
\begin{equation}
P_{no-count}(t,t+\tau) = \sum_{n=0}^{\infty}e^{-\lambda n \tau} p_n(t). \label{qed_ncp}
\end{equation}
Note that for infinitesimal time interval $\tau = \ud t$ 
one obtains
$\sum_{n=0}^{\infty}e^{-\lambda n \ud t} p_n(t)$ 
= $\sum_{n=0}^{\infty}(1-\lambda n \ud t) p_n(t)$
= $1 - \lambda \bar{n}(t)\ud t$ i.e. the sum of the one-count and no-count 
probabilities is unity.

The change in the expectation value of the number of photons
in a small time interval $\ud t$ is obtained from equation (\ref{lf_ime4})
\begin{eqnarray}
	\nonumber
\frac{\ud \bar{n}(t)}{\ud t} &=& \sum_{n=0}^{\infty} n \frac{\ud p_n(t)}{\ud t} = \lambda \sum_{n=0}^{\infty}n \left( -n p_n(t) + (n+1) p_{n+1}(t) \right)  \\
 &=& - \lambda \sum_{n=0}^{\infty}n p_n(t) = - \lambda \bar{n}(t)
\end{eqnarray}
which gives
\begin{equation}
\bar{n}(t) = \bar{n}(0)e^{-\lambda t}.
\end{equation}
Thus the expectation value of the number of photons decays exponentially in time.

Note the difference between our model and the model
of Imato \textit{et al} \cite{IUO}. 
By using the perturbation theory, they calculated how an atom
perturbs the cavity field during an infinitesimal interaction time
and thereby dissipates the photons from the cavity.
In contrast we have derived, starting from the
time-dependent Schr\"odinger equation, a master equation
for a system where the cavity field is coupled
with a detector which relaxate to a infinite reservoir.
However, both models leads to the SD model when the detector
states are reduced.

\section{Comparison of photon counting models}

The SD and the E models share four main postulates:
\\(1) 
The absorption of photons the detector takes place as instantaneous events represented by
the one-count superoperator 
$\hat{J_A}\hat{\rho}_f(t)$ = $\gamma_A \hat{A}\hat{\rho}_f(t)\hat{A}^{\dag}$
($\gamma_A$ is defined in postulate 4).
The one-count operator is a quantum jump superoperator i.e. the density operator jumps
from $\hat{\rho}_f(t)$ to $\gamma_A \hat{A}\hat{\rho}_f(t)\hat{A}^{\dag}$ in infinitesimal
time interval $[t,t+\ud t]$.
In the SD model $\hat{A} \equiv \ann{a}{}$.
The E model is obtained from the SD model by replacing the well known bosonic annihilation $\ann{a}{}$ 
and creation $\cre{a}{}$ operators by the normalized operators   
$\hat{A} \equiv$ $( \cre{a}{}\ann{a}{}+1 )^{-1/2}\ann{a}{}$ (denoted by $\hat{E}$ below)
and  $\hat{A}^{\dag} \equiv$ $\cre{a}{}( \cre{a}{}\ann{a}{}+1 )^{-1/2}$ (denoted by $\hat{E}^{\dag}$ below), respectively \cite{OMD}.
These normalized operators obey the relations $\hat{E} \ket{0}$ = 0, $\hat{E} \ket{n > 0}$ = $\ket{n-1}$ and
$\hat{E}^{\dag}\ket{n}$ = $\ket{n+1}$. 
The probability of the one-count at $[t, t+\ud t]$ is $\gamma_A$Trace$\{ \hat{A}\hat{\rho}_f\hat{A}^{\dag} \}\ud t$.
\\(2) 
Between the counts the density operator evolves according to the no-count superoperator
$\hat{S}_{\tau}\hat{\rho}_f$ = $e^{\hat{Y_A}\tau/\hbar} \hat{\rho}_f(t) e^{\hat{Y_A}^{\dag}\tau/\hbar}$,
where $\hat{Y}_A$  =$ -\iim \hat{H}_0 - \frac{1}{2}\hbar\gamma_A\hat{A}^{\dag}\hat{A}$
and $\hat{H}_0$ = $\hbar\omega \cre{a}{}\ann{a}{}$. 
Here $\tau$ is not necessarily differential.
\\(3) 
After measuring an event corresponding to the operator $\hat{\Op{O}}$ 
($\hat{\Op{O}}$ = $\hat{J}$ or $\hat{S}$),  the density operator is 
$ \hat{\rho}_f(t^{+}) = \hat{\Op{O}}\hat{\rho}_f(t) / \mathrm{Trace}\{ \hat{\Op{O}}\hat{\rho}_f(t) \}$.
\\(4) 
Furthermore, the coupling between the detector and the field is parameterized using 
a model dependent coupling coefficient $\gamma$.
The $\gamma_{sd}$ and $\gamma_e$ are not necessary equal but if we equate
the absorption rates from one photon Fock states (i.e. $\ket{1}$) we obtain
that $\gamma_{sd}$ = $\gamma_e$.

The one-count operator (Eq. (\ref{qed_1co})) and the probability (Eq. (\ref{qed_1cp})) and 
the no-count operator (Eq. (\ref{qed_nco})) and the probability (Eq. (\ref{qed_ncp})) given by our model
are equivalent with those given by the SD model (see App. \ref{sec:1cp} and \ref{sec:ncp}).
Thus our model gives an ab-initio derivation for the initially postulated SD model.

Table \ref{mnf} summarizes the expectation values of the number of photons after the one-count event,
reported previously in references \cite{DMD05,UIO}. For detailed derivations see Appendix \ref{sec:1ce}.
Also the probabilities of observing the vacuum state after the one-count event are given. 
Note the difference between the models for the thermal and coherent fields.
Our model agrees for all calculated
expectation values with results given by the SD-form of superoperators while they disagree 
with the results given by the E model as seen in Table \ref{mnf}. 

\begin{table}[!h]
\caption{The expectation values of the number of photons after a one-count event are given in the upper part. 
         In the lower part the probabilities of the vacuum state after one-count event are given
        for the thermal and the coherent fields ($p_0$ and $p_1$ are the probabilities of the zero and one photon number states, respectively). 
         If the field is initially in the Fock state $\ket{1}$ the probability $p_0(t^{+})$ = 1 after the one-count event,
        otherwise $p_0(t^{+})$ = 0.} \label{mnf}
\begin{center}
\begin{tabular}{l c r @{=} l c r @{=} l}
\hline
Initial state  & ~~ & \multicolumn{2}{c}{SD model}   & ~~ & \multicolumn{2}{c}{E model}  \\
\hline
Fock  
                & ~~  & $\bar{n}(t^{+})$ & $\bar{n}(t)-1$ & ~~  & $\bar{n}(t^{+})$ & $\bar{n}(t)-1$ \\
Thermal    & ~~  & $\bar{n}(t^{+})$ & $2\bar{n}(t)$  & ~~  & $\bar{n}(t^{+})$ & $\bar{n}(t)$ \\
Coherent   & ~~  & $\bar{n}(t^{+})$ & $\bar{n}(t)$   & ~~  & $\bar{n}(t^{+})$ & $\frac{\bar{n}(t)}{1-e^{-\bar{n}(t)}}-1$ \\
\hline
Thermal   & ~~  & $p_0(t^{+})$ & $\frac{1}{(1+\bar{n}(t))^2}$   & ~~  & $p_0(t^{+})$ & $\frac{1}{1+\bar{n}(t)}$  \\
Coherent  & ~~  & $p_0(t^{+})$ & $e^{-\bar{n}(t)}$              & ~~  & $p_0(t^{+})$ & $\frac{\bar{n}(t)}{e^{\bar{n}(t)}-1}$\\
\end{tabular}
\end{center}
\end{table}

Table \ref{coh_tab} gives photon count rates for the SD and the E models. 
The results in Table \ref{coh_tab} give the probability $w^{(1)}(t) \ud t$
of counting one photon at $[t,t+\ud t]$ and  the conditional probability $w^{(1)}(t^{+}|t) \ud t$
of counting one photon at $[t,t+\ud t]$ and the second photon at $[t^{+}, t^{+}+\ud t]$ 
($t^{+}$ = $t + \ud t$ is infinitesimally greater than $t$ 
and  $\ud t$ is the infinitesimal time that the one-count process takes
i.e. we are calculating the probability of absorbing the second photon
immediately after the first photon). 
Furthermore, the second order coherence degree
\begin{equation}
g^{(2)}(t, t^{+}) =  \frac{ w^{(1)}(t^{+}|t) }{ w^{(1)}(t^{+}) }
\end{equation}
is also given in Table \ref{coh_tab}. 
For the derivation of the photon correlation function see Appendix \ref{sec:g2}.
Our  results agree with the results of the SD model.

\begin{table}[!h]
\caption{The one-count rates ($w^{(1)}(t)$), conditional one-count rates ($w^{(1)}(t^{+}|t)$), 
and the second order coherence degrees ($g^{(2)}(t,t^{+})$) given by the SD and the E models. } \label{coh_tab}
\begin{center}
\begin{tabular}{l | c c | c c | c  c}
\hline
Initial state   & $w_{sd}^{(1)}(t)$  & $w_{sd}^{(1)}(t^{+}|t)$ & $w_{e}^{(1)}(t)$  & $w_{e}^{(1)}(t^{+}|t)$ & $g^{(2)}_{sd}(t,t^{+})$ & $g^{(2)}_{e}(t,t^{+})$   \\
\hline
Fock       &  $\gamma_{sd} \bar{n}(t)$  &  $\gamma_{sd} (\bar{n}(t)-1)$ &  $\gamma_{e}$   &  $\gamma_{e}$ &  $\frac{\bar{n}(t)-1}{\bar{n}(t)}$    & 1 \\
Thermal    &  $\gamma_{sd} \bar{n}(t)$  &  $2 \gamma_{sd} \bar{n}(t)$   &  $\gamma_{e} \frac{\bar{n}(t)}{1+\bar{n}(t)}$   &  $\gamma_{e} \frac{\bar{n}(t)}{1+\bar{n}(t)}$  & 2  & 1 \\
Coherent   &  $\gamma_{sd} \bar{n}(t)$  &  $\gamma_{sd} \bar{n}(t)$ &  $\gamma_{e} (1-e^{-\bar{n}(t)})$    &  $\gamma_{e} \left(1-\frac{\bar{n}(t)}{e^{\bar{n}(t)}-1}\right)$  & 1  & $\frac{e^{\bar{n}(t)}-(\bar{n}(t)+1)}{e^{\bar{n}(t)}+e^{-\bar{n}(t)}-2}$ 
\end{tabular}
\end{center}
\end{table}

The evolution of the expectation value of the number of photons
in the SD and E models are respectively (see  Appendix \ref{sec:empn})
\begin{eqnarray}
\bar{n}_{sd}(t) &=& \bar{n}(0)e^{-\gamma_{sd} t} \\
\bar{n}_e(t) &=& \bar{n}(0) + \gamma_{e} \int_{0}^{t} (p_0(t')-1)\ud t'. 
\end{eqnarray}
Thus $\bar{n}(t)$ decays exponentially in the SD model irrespective of the field type
while in the E model $\bar{n}(t)$ depends on the time integral of the
vacuum state probability which is different for different field types.
Again our  results agree with the results of the SD model.


\hspace{2cm}



%
\section{Discussion}
%


The E model of photon counting was originally motivated by arguing that the results
given by the SD model are unphysical. Next we will show
by selected examples that, on the contrary, it is the SD model that gives 
physical and also intuitionally right answers.

In references \cite{DMD05, DMD06, DMD07} Dodonov \textit{et al.} 
gave the following main arguments in favor of the E model: 
(1) The expectation value of the number of photons
may increase when operating with the SD one-count operator \cite{UIO}, see Table \ref{mnf}. 
(2) The absorption rate of photons in the SD model
is proportional to the number of photons and does not saturate for high $\bar{n}$.

\subsubsection{Increase of $\bar{n}$ after one-count event}

In the SD model the expectation value of the number of 
photons after the observed absorption of one photon (i.e. after the one-count event)
may be greater than before the absorption. 
For example after detecting of a photon with the SD one-count 
operator from the thermal field  the expectation value
of the number of photons doubles (see Table \ref{mnf}). 
The assumption that this feature of the SD model is 
unphysical was the main justification for the E model. 
This behavior caused by the measurement back action 
to the cavity field is, however, in agreement with the photon bunching effect:
When a photon arrives to a detector the expectation value of the number of photons
raises and thus it is more probable to detect another photon.
The growth of the  expectation value of the number of photons
is not unphysical and  means that the states which had one or more photons
become more probable after the  absorption of the first photon. 

Furthermore, we have shown that in our model (as well as in the SD model)
the expectation value of the number of the photons has exponential decay in time
for all fields as it should. Thus, even if the expectation value of the number 
of photons may increase in the one-count event, on average the expectation
value of the number of the photons decreases on every time interval
during which no measurement induced projection of the field state occurs.

\subsubsection{Saturation of absorption rate}

The one-count rates for the SD and E models are
$w_{sd}^{(1)}(t)$ = $\gamma_{sd}\bar{n}(t)$ and $w_{e}^{(1)}(t)$ = $\gamma_{e}(1-p_0(t))$, respectively.
The fundamental difference between the models is that
in the SD model the photocount rate  is proportional to the  expectation value of the number of photons
while in the E model it is proportional to the probability that photons exists. 
Thus the absorption probability saturates in the E model. This may be a reasonable
assumption for the very high intensities when the detector itself may saturate.
However, the possible saturation of absorption probability is a internal property of the
detector and should be an auxiliary model property,
not a built in property of the quantum jump superoperators. 
In principle it is possible that at high intensities the detector could
scatter some photons back to the cavity.
This would lead to non-constant $\gamma$ which
the present models (SD or E) can not adopt as such.

Let us consider two simple  examples: 
(a) The field is in the state $\ket{\Psi}$ = $(\ket{0} + \ket{1})/\sqrt{2}$. 
The one-count rates are
$w_{sd}^{(1)}(t)$ = $\gamma_{sd}/2$ 
and
$w_{e}^{(1)}(t)$ = $\gamma_{e}/2$.
(b) The field is the state $\ket{\Psi}$ = $(\ket{0} + \ket{100})/\sqrt{2}$. 
The one-count rates are
$w_{sd}^{(1)}(t)$ = $100\gamma_{sd}/2$ 
and
$w_{e}^{(1)}(t)$ = $\gamma_{e}/2$.
The behavior of the SD model seems more reasonable because it should
be more probable to detect photons when the expectation value of the photons
is larger.

\subsubsection{Second order coherence degree}

The photon bunching effect is related to the  second order coherence: 
If $0 \le g^{(2)}(0) < 1$ the light is antibunched, if $g^{(2)}(0) = 1$ the light is
nonbunched or random, and if $g^{(2)}(0) > 1$ the light is bunched \cite{QTL}.
Here the argument zero means that the correlation is measured with zero time delay. 
Examples of light fields obeying the above conditions are the Fock state, 
the coherent state and the thermal state, respectively.
The calculated second order coherence degrees  are given in Table  \ref{coh_tab}.

For single-mode fields the second order coherence degrees are independent
of the time delay between the measurements. The values are
$g^{(2)}=1$ for the coherent field,  $g^{(2)}=2$ for the thermal field
and $g^{(2)}=m/(m-1)$ for the Fock state $\ket{m}$ ($m \ge 2$)  \cite{QTL}. 
Comparing the results given in Table \ref{coh_tab} we note that the SD model reproduces 
the second order coherence degrees of the three standard example fields.

The results given by our model (as well as the SD model) are equal with the experimental
results given in Reference \cite{LLZ} for thermal and coherent fields.
Note, however, that the measurement results in Reference \cite{LLZ} are 
not given for light fields in a closed cavity but for continuous wave laser 
with a Gaussian frequency distribution for thermal light.
We are comparing the results measured with zero time delay since
the zero time delay cancels the multi-mode effect. Furthermore, we expect 
that the measurement results of the second order coherence degree
in free space correspond to our theoretical results since $g^{(2)}$  depends only on 
the statistics of the  light field, not the spatial distribution.

Dodonov \textit{et al.} \cite{DMD05} also concluded that 
the SD model always predicts the photon bunching phenomenon for any 
initial field. We have shown (see Table \ref{coh_tab}) that this conclusion  
given in reference \cite{DMD05} is incorrect.

\section{Conclusions}

In this paper 
we have shown that by starting from 
the time-dependent Schr\"odinger equation 
one can derive an exact difference equation for the density operator $\hat{\rho}_I$
of the combined detector-field system.
Reducing this change of the density operator with respect to the detector
variables gives a time integrable master equation for the field 
density operator. This gives the first principle expressions (Eqs. (\ref{qed_1co}) and (\ref{qed_nco}))
for the one-count and no-count operators. Our results for the one-count and no-count
operators agree with the SD model (see App. \ref{sec:1cp} and \ref{sec:ncp} for comparison)
which therefore, in spite of being initially introduced
ad hoc \cite{SD}, is actually an exact quantum mechanical result.


%

\appendix

\section{Derivations}

\subsection{Derivation of master equation} \label{sec:qed_der}

The time-dependent Schr\"odinger equation is
\begin{equation}
\iim\hbar \frac{\ud}{\ud t}\ket{\Psi_s(t)} = ( \hat{H}_0^s + \hat{H}_I^s )\ket{\Psi_s(t)},
\end{equation}
where $\hat{H}_0^s$ 
= $\hbar\omega_0\cre{a}{}\ann{a}{}$ + $\hbar\omega_A \ket{e}\bra{e}$
is the time-independent field + detector Hamiltonian and 
$\hat{H}_I^s$ = $\hbar\Omega \left( \ann{a}{}\ket{e}\bra{g} + \cre{a}{}\ket{g}\bra{e} \right)$
is the time-independent interaction Hamiltonian.
The interaction representation is given by unitary transformation
\begin{equation}
\ket{\Psi_I(t)} = e^{\iim \hat{H}_0^s t/\hbar}\ket{\Psi_s(t)},
\end{equation}
where $\ket{\Psi_I(0)}$ = $\ket{\Psi_s(0)}$.
The Schr\"odinger equation gives 
\begin{eqnarray}
	\nonumber
\iim\hbar \frac{\ud}{\ud t}\ket{\Psi_I(t)} &=& \iim\hbar \left( \frac{\iim \hat{H}_0^s }{\hbar} \ket{\Psi_I(t)} +   e^{\iim \hat{H}_0^s t/\hbar} \frac{\ud}{\ud t}\ket{\Psi_s(t)} \right) \\
	\nonumber
&=&  -\hat{H}_0^s \ket{\Psi_I(t)}  +  e^{\iim \hat{H}_0^s t/\hbar}( \hat{H}_0^s + \hat{H}_I^s )e^{-\iim \hat{H}_0^s t/\hbar} \ket{\Psi_I(t)}  \\
	\nonumber
&=&   e^{\iim \hat{H}_0^s t/\hbar} \hat{H}_I^s e^{-\iim \hat{H}_0^s t/\hbar} \ket{\Psi_I(t)}\\
&=&    \hat{H}_I^s  \ket{\Psi_I(t)}, \label{SE_int}
\end{eqnarray}
where the last form holds, since $[\hat{H}_0^s, \hat{H}_I^s]=0$. Note that, since
$\hat{H}_0^s$ and $\hat{H}_I^s$ commute, the interaction Hamiltonian
in the interaction picture is same as in the Schr\"odinger picture.
Equation (\ref{SE_int}) gives following differential equation
for the density operator of the coupled field-detector system
in the interaction picture
\begin{equation}
\iim \hbar \frac{\ud \hat{\rho}_I(t)}{\ud t} = [\hat{H}_I^s, \hat{\rho}_I(t)]. \label{tddo}
\end{equation}
The density operator can be written using the Taylor series as
\begin{eqnarray}
\hat{\rho}_I(dt) &=& \hat{\rho}_I(0) + \left. \frac{\ud \hat{\rho}_I(t)}{\ud t}\right|_{t=0} \ud t 
  + \frac{1}{2}\left. \frac{\ud^2 \hat{\rho}_I(t)}{\ud t^2}\right|_{t=0} (\ud t)^2 + \dots.
\end{eqnarray}
Since the interaction Hamiltonian is time independent, equation (\ref{tddo}) can be used to obtain
the higher derivatives in the Taylor series
\begin{eqnarray}
\frac{\ud \hat{\rho}_I(t)}{\ud t} &=&  -\frac{\iim}{\hbar}[\hat{H}_I^s, \hat{\rho}_I(t)] \\
	\nonumber
\frac{\ud^2 \hat{\rho}_I(t)}{\ud t^2} &=& -\frac{\iim}{\hbar}[\hat{H}_I^s, \frac{\ud \hat{\rho}_I(t)}{\ud t}] \\
          &=& -\frac{1}{\hbar^2}\left[\hat{H}_I^s, [\hat{H}_I^s, \hat{\rho}_I(t)]\right].
\end{eqnarray}
Taking the terms up to second order in $\ud t$ the Taylor expansion can be written as
\begin{eqnarray}
\hat{\rho}(dt) &=& \hat{\rho}_I(0)  -\frac{\iim}{\hbar}[\hat{H}_I^s, \hat{\rho}_I(0)] \ud t
  - \frac{1}{2\hbar^2}\left[\hat{H}_I^s, [\hat{H}_I^s, \hat{\rho}_I(0)]\right] (\ud t)^2,
\end{eqnarray}
where the double commutator is
$\left[\hat{H}_I^s, [\hat{H}_I^s, \hat{\rho}_I(0)]\right]$ = 
$[\hat{H}_I^s, \hat{H}_I^s\hat{\rho}_I(0)-\hat{\rho}_I(0)\hat{H}_I^s ]$ =
$\hat{H}_I^s\hat{H}_I^s\hat{\rho}_I(0)+\hat{\rho}_I(0)\hat{H}_I^s\hat{H}_I^s-2\hat{H}_I^s\hat{\rho}(0)\hat{H}_I^s $
giving
\begin{eqnarray}
\hat{\rho}(dt) &=& \hat{\rho}_I(0)  -\frac{\iim}{\hbar}[\hat{H}_I^s, \hat{\rho}_I(0)] \ud t
  + \frac{1}{2\hbar^2}\left(2\hat{H}_I^s\hat{\rho}(0)\hat{H}_I^s - \{ \hat{H}_I^s\hat{H}_I^s, \hat{\rho}_I(0) \}\right).
\end{eqnarray}

\subsection{One-count probabilities and rates} \label{sec:1cp}

The one-count probability in the time interval $[t,t+\ud t]$
is given by  $\gamma_A$Trace$\{\hat{A}^{\dag}\hat{A}\hat{\rho}_f\}\ud t$.
Thus $\gamma_A$Trace$\{\hat{A}^{\dag}\hat{A}\hat{\rho}_f\}$ is the photon count rate.
For the SD and E models we, respectively, obtain the photon count rates 
( Trace$\{ \cdot \}$ = $\sum_{n=0}^{\infty}\bra{n} \cdot \ket{n}$)
\begin{eqnarray}
w_{sd}(t) 
&=& \gamma_{sd} \sum_{n=0}^{\infty} n p_n(t) = \gamma_{sd} \bar{n}(t) \\
w_{e}(t) 
&=& \gamma_{e} \sum_{n=1}^{\infty} p_n(t) = \gamma_{e}(1 - p_0(t))
\end{eqnarray}
The one-count probabilities are obtained by multiplying the rates with $\ud t$. Thus
$P_{sd}^{one-count}(t)$ = $\gamma_{sd} \bar{n}(t) \ud t$
and
$P_{e}^{one-count}(t)$ = $\gamma_{e}(1 - p_0(t)) \ud t$.

\subsection{One-count event} \label{sec:1ce}

If the one-count event is detected the density operator must change
in accordance to the operation by the one-count operator and normalization.
The density operator after one-count is
$\hat{\rho}_f(t^{+}) = \frac{ \hat{A}\hat{\rho}_f(t)\hat{A}^{\dag} }{\mathrm{Trace}\{ \hat{A}\hat{\rho}_f(t)\hat{A}^{\dag} \}}$
= $ \frac{ \hat{A}\hat{\rho}_f(t)\hat{A}^{\dag} }{\mathrm{Trace}\{ \hat{A}^{\dag}\hat{A}\hat{\rho}_f(t) \}} $.
For the SD and E models we, respectively, obtain 
\begin{eqnarray}
	\nonumber
\hat{\rho}_f^{sd}(t^{+}) &=& \frac{1}{\bar{n}(t)}\sum_{n,n'=0}^{\infty}p_{n,n'}(t)\sqrt{nn'}\ket{n-1}\bra{n'-1} \\
       &=& \frac{1}{\bar{n}(t)}\sum_{n=0}^{\infty}p_{n+1, n'+1}(t)\sqrt{(n+1)(n'+1)}\ket{n}\bra{n'} \\
\hat{\rho}_f^{e}(t^{+}) &=& \frac{1}{1-p_0(t)}\sum_{n,n'=1}^{\infty}p_{n,n'}(t)\ket{n-1}\bra{n'-1} = \frac{\sum_{n,n'=0}^{\infty}p_{n+1,n'+1}(t)\ket{n}\bra{n'}}{1-p_0(t)}.
\end{eqnarray}
Thus the new probabilities of the field states of $n$ photons are
\begin{eqnarray}
p_n^{sd}(t^{+}) &=& \frac{n+1}{\bar{n}(t)}p_{n+1}(t) \\
p_n^{e}(t^{+}) &=& \frac{p_{n+1}(t)}{1-p_{0}(t)}.
\end{eqnarray}

\subsection{No-count probability }\label{sec:ncp}

The no-count operator is 
$\hat{S}_{\tau}\hat{\rho}_f = e^{\hat{Y_A}\tau/\hbar} \hat{\rho}_f e^{\hat{Y_A}^{\dag}\tau/\hbar}$,
where $\hat{Y}_A = -\iim \hat{H}_0 - \frac{1}{2}\gamma_A\hbar\hat{A}^{\dag}\hat{A}$
and $\hat{H}_0 = \hbar\omega_0 \cre{a}{}\ann{a}{}$.
The series expansion in the SD model is

\begin{equation}
e^{ \left(  -\iim \omega_0 - \frac{1}{2}\gamma \right) \cre{a}{}\ann{a}{}  \tau} \ket{m}
= \sum_{n=0}^{\infty}\frac{(-\iim \omega_0 - \frac{1}{2}\gamma)^n (\cre{a}{}\ann{a}{})^n \tau^n  }{n!} \ket{m}
=  e^{ \left(  -\iim \omega_0 - \frac{1}{2}\gamma \right) m \tau} \ket{m}
\end{equation}
Thus the time evolution of the density operator in the no-count event is
\begin{eqnarray}
 	\nonumber
 e^{\hat{Y}\tau} \hat{\rho}_f e^{\hat{Y}^{\dag}\tau} 
 &=& \sum_{m,m'=0}^{\infty} e^{-\iim\omega_0(m-m')\tau   -\gamma\frac{m+m'}{2}\tau} p_{m,m'}\ket{m}\bra{m'} .
\end{eqnarray}
The series expansion in the E model is
\begin{equation}
e^{\hat{Y}\tau} = e^{ -\iim \omega_0\cre{a}{}\ann{a}{}\tau - \frac{1}{2}\gamma \hat{E}^{\dag}\hat{E}\tau} 
=     \sum_{n=0}^{\infty}\frac{(-\iim \omega_0)^n (\cre{a}{}\ann{a}{})^n \tau^n  }{n!} 
        \sum_{k=0}^{\infty}\frac{(-\frac{1}{2}\gamma)^k ( \hat{E}^{\dag}\hat{E} )^k \tau^k  }{k!}
\end{equation}
giving
\begin{equation}
e^{\hat{Y}\tau}\ket{m} 
=  e^{-\iim \omega_0 m \tau} \left\{
\begin{array}{r}
     \sum_{k=0}^{\infty}\frac{(-\frac{1}{2}\gamma)^k  \tau^k  }{k!} \ket{m} \\
     \ket{0}
\end{array} \right.
= \left\{
\begin{array}{rl}
     e^{-\iim \omega_0 m \tau -\frac{1}{2}\gamma\tau } \ket{m}, & m > 0 \\
     \ket{0}, & m = 0
\end{array} \right.
\end{equation}
and the evolution of the density operator
\begin{eqnarray}
 	\nonumber
 e^{\hat{Y}\tau} \hat{\rho}_f e^{\hat{Y}^{\dag}\tau} &=& 
 \sum_{m,m'=0}^{\infty} e^{ -\iim \omega_0\cre{a}{}\ann{a}{}\tau - \frac{1}{2}\gamma \hat{E}^{\dag}\hat{E}\tau} p_{m,m'}\ket{m}\bra{m'} e^{ +\iim \omega_0\cre{a}{}\ann{a}{}\tau - \frac{1}{2}\gamma \hat{E}^{\dag}\hat{E}\tau} \\
 	\nonumber
  &=& p_{0,0}\ket{0}\bra{0} + \sum_{m=1}^{\infty}\left[p_{0,m}\ket{0}\bra{m}e^{+\iim\omega_0m\tau - \frac{1}{2}\gamma\tau} + p_{m,0}\ket{m}\bra{0}e^{-\iim\omega_0m\tau - \frac{1}{2}\gamma\tau}\right] \\ 
  && + \sum_{m,m'=1}^{\infty}p_{m,m'}e^{-\iim\omega_0(m-m')\tau - \gamma\tau}\ket{m}\bra{m'}.
\end{eqnarray}
We are interested only on the diagonal elements to be able to calculate the evolution of the probability of
the $n$ photon state by taking the matrix elements $\me{n}{\cdot}{n}$.
Thus the SD model gives
\begin{equation}
 \me{n}{ \hat{S}_{\tau}\hat{\rho}_f(0) }{n} = e^{-n \gamma_{sd} \tau}p_n(0) \label{sp_sd_nc}
\end{equation}
and the E model gives
\begin{equation}
\me{n}{ \hat{S}_{\tau}\hat{\rho}_f(0) }{n} = \left\{ \begin{array}{ll} e^{-\gamma_{e} \tau}p_n(0), & n>0 \\ p_0(0), & n=0 \end{array} \right. 
         \label{sp_e_nc}         
\end{equation}
The no-count probabilities at the time interval $[t,t+\tau]$ are the sums over the probabilities $\me{n}{ \hat{S}_{\tau}\hat{\rho}_f(t) }{n}$.  
The SD model and the E model, respectively, give
\begin{eqnarray}
P_{sd}^{no-count}(t,t+\tau) &=& \sum_{n=0}^{\infty} e^{-n \gamma_{sd} \tau} p_n(t) \\
P_{e}^{no-count}(t,t+\tau)  &=& p_0(t)+(1-p_0(t))e^{-\gamma_{e} \tau}.
\end{eqnarray}

\subsection{No-count event} \label{sec:nce}

If we know that the one-count event  has not happened,  
the density operator must change in accordance to the operation
by the no-count operator and normalization. Thus the probabilities
of $n$ photon states are obtained by normalizing 
equations (\ref{sp_sd_nc}) and (\ref{sp_e_nc})
giving for the SD and E models, respectively,
\begin{eqnarray}
 p_n(t+\tau) &=& \frac{e^{-n \gamma_{sd} \tau}p_n(t)}{ \sum_{n=0}^{\infty} e^{-n \gamma_{sd} \tau}p_n(t) }\\
 p_n(t+\tau) &=& \left\{ \begin{array}{ll} \frac{e^{-\gamma_{e} \tau}p_n(t)}{ p_0(t)+(1-p_0(t))e^{-\gamma_{e} \tau} }, & n>0 
         \\ \frac{p_0(t)}{p_0(t)+(1-p_0(t))e^{-\gamma_{e} \tau}}, & n=0. \end{array} \right. 
\end{eqnarray}

\subsection{Evolution of expectation value of photon number} \label{sec:empn}

The density matrix evolves according to equation \cite{SD,DMD05,DMD07}
\begin{eqnarray}
\frac{\ud \hat{\rho}}{\ud t} 
&=& -\iim \omega \left(\cre{a}{}\ann{a}{}\hat{\rho} - \hat{\rho}\cre{a}{}\ann{a}{}\right)
	 +\left( \gamma_A \hat{A}\hat{\rho}\hat{A}^{\dag} - \frac{\gamma_A}{2}(\hat{A}^{\dag}\hat{A}\hat{\rho} + \hat{\rho}\hat{A}^{\dag}\hat{A} ) \right).
\end{eqnarray}
The probabilities of $n$ photon states are given by the diagonal elements $\me{n}{\cdot}{n}$.
Thus we obtain for the SD model 
\begin{eqnarray}
\frac{\ud p_{n}(t)}{\ud t} &=&  \gamma_{sd}\left(   (n+1) p_{n+1}(t)  - p_{n}(t) n \right)	
\end{eqnarray}
and for the E model
\begin{eqnarray}
\frac{\ud p_{n \ge 1}(t)}{\ud t} &=&  \gamma_{e}\left(    p_{n+1}(t)  - p_{n}(t) \right) \\	
\frac{\ud p_{0}(t)}{\ud t} &=&  \gamma_{e}    p_{1}(t). 	
\end{eqnarray}

The expectation value of the number of photons is defined as
\begin{eqnarray}
\bar{n}(t) = \sum_{n=0}^{\infty}np_n(t).
\end{eqnarray}
For the SD model the time derivation of $\bar{n}(t)$ is given by
\begin{eqnarray}
	\nonumber
\frac{\ud \bar{n}(t)}{\ud t} &=& \sum_{n=0}^{\infty}n\frac{ \ud p_n(t)}{\ud t} 
	= \gamma_{sd} \sum_{n=0}^{\infty} \left[ (n+1)n p_{n+1}(t) - n^2 p_{n}(t) \right] \\
	\nonumber
&=& \gamma_{sd} \left( \sum_{n=0}^{\infty} (n+1)n p_{n+1}(t) - \sum_{n=0}^{\infty} n^2 p_{n}(t) \right) 
           = \gamma_{sd}\left( \sum_{n=1}^{\infty} n(n-1) p_{n}(t) - \sum_{n=1}^{\infty} n^2 p_{n}(t) \right)  \\
	\nonumber
 &=& -\gamma_{sd} \sum_{n=0}^{\infty}n p_n(t) = -\gamma_{sd} \bar{n}(t).
\end{eqnarray}
This has a solution
\begin{equation}
\bar{n}(t) = \bar{n}(0)e^{-\gamma_{sd} t}. 
\end{equation}
For the E model we correspondingly obtain
\begin{eqnarray}
	\nonumber
\frac{\ud \bar{n}(t)}{\ud t} &=& \sum_{n=0}^{\infty}n\frac{ \ud p_n(t)}{\ud t} 
	= \gamma_{e} \sum_{n=0}^{\infty} \left[ n p_{n+1}(t) - n p_{n}(t) \right] \\
	\nonumber
&=& \gamma_{e}\left( \sum_{n=0}^{\infty} n p_{n+1}(t) - \sum_{n=0}^{\infty}  n p_{n}(t) \right) 
    = \gamma_{e}\left( \sum_{n=1}^{\infty} (n-1) p_{n}(t) - \sum_{n=1}^{\infty}  n p_{n}(t) \right) \\
	\nonumber
&=& -\gamma_{e} \sum_{n=1}^{\infty}p_n(t) = -\gamma_{e} (1-p_0(t)).
\end{eqnarray}
Time integration gives for the expectation value of the number of photons a solution
\begin{equation}
\bar{n}(t) = \bar{n}(0) + \gamma_{e} \int_{0}^{t} (p_0(t')-1)\ud t'. 
\end{equation}

\subsection{Second order coherence degree} \label{sec:g2}

The second order coherence degree is \cite{QTL,QTOC}
\begin{equation}
g^{(2)}(\V{r}_1, t_1, \V{r}_2, t_2, \V{r}_2, t_2 ,\V{r}_1, t_1) =
\frac{G^{(2)}(\V{r}_1, t_1, \V{r}_2, t_2, \V{r}_2, t_2 ,\V{r}_1, t_1)}{G^{(1)}(\V{r}_1, t_1, \V{r}_1, t_1) G^{(1)}(\V{r}_2, t_2, \V{r}_2, t_2)},
\end{equation}
where 
\begin{eqnarray}
	\nonumber
&& G^{(2)}(\V{r}_1, t_1, \V{r}_2, t_2, \V{r}_2, t_2 ,\V{r}_1, t_1)   \\
	\nonumber
&=&  \mathrm{Tr}\{ \hat{\rho}_f \hat{\V{E}}^{(-)}(\V{r}_1,t_1)  \hat{\V{E}}^{(-)}(\V{r}_2,t_2) \hat{\V{E}}^{(+)}(\V{r}_{2},t_{2}) \dots \hat{\V{E}}^{(+)}(\V{r}_{1},t_{1}) \}
\end{eqnarray}
with $\hat{\V{E}}^{(-)}(\V{r},t)$ and $\hat{\V{E}}^{(+)}(\V{r},t)$ being the negative and positive frequency parts of the electric field operator. 
The two-fold delayed coincidence rate i.e. the counting rate per (unit time)$^2$ is given by \cite{QTOC}
\begin{equation}
w^{(2)}(\V{r}_1, t_1, \V{r}_2, t_2, \V{r}_2, t_2, \V{r}_1, t_1)
= s^2 G^{(2)}(\V{r}_1, t_1, \V{r}_2, t_2, \V{r}_2, t_2, \V{r}_1, t_1),
\end{equation}
where $s$ is the sensitivity of the detector. We consider only the temporal correlation
so we assume that all of the position vectors are equal and drop the spatial coordinate.
We can now use well know formula of conditional
probability: the probability that an event $B$ occurs with the condition that $A$ has happened 
is $p(B|A) = p(B \cap A)/p(A)$. Thus $p(B \cap A) = p(B|A)p(A)$ giving
$w^{(2)}(t^{+},t)(\ud t)^2$ = $w^{(1)}(t^{+}|t) \ud t$ $w^{(1)}(t)\ud t$, 
where we are considering correlation with infinitesimal time difference.
Furthermore, we can write the second order coherence degree using the count rates 
\begin{eqnarray}
g^{(2)}(t, t^{+}) &=& \frac{w^{(2)}(t, t^{+})}{w^{(1)}(t) ~  w^{(1)}(t^{+})} 
=\frac{ w^{(1)}(t^{+}|t) w^{(1)}(t) }{ w^{(1)}(t)  w^{(1)}(t^{+})} = \frac{ w^{(1)}(t^{+}|t) }{ w^{(1)}(t^{+}) }, \label{coh_form}
\end{eqnarray}
where we, furthermore, assume that $w^{(1)}(t^{+})$ = $w^{(1)}(t)$ due to the differential time difference.


\bibliography{SDE_viitteet}

\begin{thebibliography}{12}
\expandafter\ifx\csname natexlab\endcsname\relax\def\natexlab#1{#1}\fi
\expandafter\ifx\csname bibnamefont\endcsname\relax
  \def\bibnamefont#1{#1}\fi
\expandafter\ifx\csname bibfnamefont\endcsname\relax
  \def\bibfnamefont#1{#1}\fi
\expandafter\ifx\csname citenamefont\endcsname\relax
  \def\citenamefont#1{#1}\fi
\expandafter\ifx\csname url\endcsname\relax
  \def\url#1{\texttt{#1}}\fi
\expandafter\ifx\csname urlprefix\endcsname\relax\def\urlprefix{URL }\fi
\providecommand{\bibinfo}[2]{#2}
\providecommand{\eprint}[2][]{\url{#2}}

\bibitem[{\citenamefont{Dodonov et~al.}(2007)\citenamefont{Dodonov, Mizrahi,
  and Dodonov}}]{DMD07}
\bibinfo{author}{\bibfnamefont{A.~V.} \bibnamefont{Dodonov}},
  \bibinfo{author}{\bibfnamefont{S.~S.} \bibnamefont{Mizrahi}},
  \bibnamefont{and} \bibinfo{author}{\bibfnamefont{V.~V.}
  \bibnamefont{Dodonov}}, \bibinfo{journal}{Physical Review A}
  \textbf{\bibinfo{volume}{75}}, \bibinfo{pages}{013806}
  (\bibinfo{year}{2007}).

\bibitem[{\citenamefont{Kelley and Kleiner}(1964)}]{KK}
\bibinfo{author}{\bibfnamefont{P.~L.} \bibnamefont{Kelley}} \bibnamefont{and}
  \bibinfo{author}{\bibfnamefont{W.~H.} \bibnamefont{Kleiner}},
  \bibinfo{journal}{Physical Review} pp. \bibinfo{pages}{316--334}
  (\bibinfo{year}{1964}).

\bibitem[{\citenamefont{Mandel}(1981)}]{M81}
\bibinfo{author}{\bibfnamefont{L.}~\bibnamefont{Mandel}},
  \bibinfo{journal}{Journal of Modern Optics} \textbf{\bibinfo{volume}{28}},
  \bibinfo{pages}{1447} (\bibinfo{year}{1981}).

\bibitem[{\citenamefont{Srinivas and Davies}(1981)}]{SD}
\bibinfo{author}{\bibfnamefont{M.~D.} \bibnamefont{Srinivas}} \bibnamefont{and}
  \bibinfo{author}{\bibfnamefont{E.~B.} \bibnamefont{Davies}},
  \bibinfo{journal}{Journal of Modern Optics} \textbf{\bibinfo{volume}{28}},
  \bibinfo{pages}{981} (\bibinfo{year}{1981}).

\bibitem[{\citenamefont{de~Oliveira et~al.}(2003)\citenamefont{de~Oliveira,
  Mizrahi, and Dodonov}}]{OMD}
\bibinfo{author}{\bibfnamefont{M.~C.} \bibnamefont{de~Oliveira}},
  \bibinfo{author}{\bibfnamefont{S.~S.} \bibnamefont{Mizrahi}},
  \bibnamefont{and} \bibinfo{author}{\bibfnamefont{V.~V.}
  \bibnamefont{Dodonov}}, \bibinfo{journal}{Journal of Optics B: Quantum and
  Semiclassical Optics} \textbf{\bibinfo{volume}{5}}, \bibinfo{pages}{S271}
  (\bibinfo{year}{2003}).

\bibitem[{\citenamefont{Dodonov et~al.}(2005)\citenamefont{Dodonov, Mizrahi,
  and Dodonov}}]{DMD05}
\bibinfo{author}{\bibfnamefont{A.~V.} \bibnamefont{Dodonov}},
  \bibinfo{author}{\bibfnamefont{S.~S.} \bibnamefont{Mizrahi}},
  \bibnamefont{and} \bibinfo{author}{\bibfnamefont{V.~V.}
  \bibnamefont{Dodonov}}, \bibinfo{journal}{Journal of Optics B: Quantum and
  Semiclassical Optics} \textbf{\bibinfo{volume}{7}}, \bibinfo{pages}{99}
  (\bibinfo{year}{2005}).

\bibitem[{\citenamefont{Dodonov et~al.}(2006)\citenamefont{Dodonov, Mizrahi,
  and Dodonov}}]{DMD06}
\bibinfo{author}{\bibfnamefont{A.~V.} \bibnamefont{Dodonov}},
  \bibinfo{author}{\bibfnamefont{S.~S.} \bibnamefont{Mizrahi}},
  \bibnamefont{and} \bibinfo{author}{\bibfnamefont{V.~V.}
  \bibnamefont{Dodonov}}, \bibinfo{journal}{Physical Review A}
  \textbf{\bibinfo{volume}{74}}, \bibinfo{pages}{033823}
  (\bibinfo{year}{2006}).

\bibitem[{\citenamefont{Li et~al.}(2007)\citenamefont{Li, Li, Zhang, Wang,
  Zhang, Wang, and Zhang}}]{LLZ}
\bibinfo{author}{\bibfnamefont{Y.}~\bibnamefont{Li}},
  \bibinfo{author}{\bibfnamefont{G.}~\bibnamefont{Li}},
  \bibinfo{author}{\bibfnamefont{Y.~C.} \bibnamefont{Zhang}},
  \bibinfo{author}{\bibfnamefont{X.~Y.} \bibnamefont{Wang}},
  \bibinfo{author}{\bibfnamefont{J.}~\bibnamefont{Zhang}},
  \bibinfo{author}{\bibfnamefont{J.~M.} \bibnamefont{Wang}}, \bibnamefont{and}
  \bibinfo{author}{\bibfnamefont{T.~C.} \bibnamefont{Zhang}},
  \bibinfo{journal}{Physical Review A} \textbf{\bibinfo{volume}{76}},
  \bibinfo{eid}{013829} (pages~\bibinfo{numpages}{5}) (\bibinfo{year}{2007}).

\bibitem[{\citenamefont{Imoto et~al.}(1990)\citenamefont{Imoto, Ueda, and
  Ogawa}}]{IUO}
\bibinfo{author}{\bibfnamefont{N.}~\bibnamefont{Imoto}},
  \bibinfo{author}{\bibfnamefont{M.}~\bibnamefont{Ueda}}, \bibnamefont{and}
  \bibinfo{author}{\bibfnamefont{T.}~\bibnamefont{Ogawa}},
  \bibinfo{journal}{Physical Review A} \textbf{\bibinfo{volume}{41}},
  \bibinfo{pages}{4127} (\bibinfo{year}{1990}).

\bibitem[{\citenamefont{Ueda et~al.}(1990)\citenamefont{Ueda, Imoto, and
  Ogawa}}]{UIO}
\bibinfo{author}{\bibfnamefont{M.}~\bibnamefont{Ueda}},
  \bibinfo{author}{\bibfnamefont{N.}~\bibnamefont{Imoto}}, \bibnamefont{and}
  \bibinfo{author}{\bibfnamefont{T.}~\bibnamefont{Ogawa}},
  \bibinfo{journal}{Physical Review A} \textbf{\bibinfo{volume}{41}},
  \bibinfo{pages}{3891} (\bibinfo{year}{1990}).

\bibitem[{\citenamefont{Loudon}(1983)}]{QTL}
\bibinfo{author}{\bibfnamefont{R.}~\bibnamefont{Loudon}},
  \emph{\bibinfo{title}{The Quantum Theory of Light}}
  (\bibinfo{publisher}{Oxford University Press}, \bibinfo{year}{1983}).

\bibitem[{\citenamefont{Glauber}(2007)}]{QTOC}
\bibinfo{author}{\bibfnamefont{R.~J.} \bibnamefont{Glauber}},
  \emph{\bibinfo{title}{Quantum Theory of Optical Coherence}}
  (\bibinfo{publisher}{Wiley-VHC}, \bibinfo{year}{2007}).

\end{thebibliography}

\end{document}